\documentclass[sigconf, nonacm]{acmart}

\usepackage{pgfplotstable}
\usepackage{pgfplots}
\pgfplotsset{compat=1.18}
\usepackage{glossaries}
\newacronym{AI}{AI}{Artificial Intelligence}
\newacronym{AIBOM}{AIBOM}{AI Bill of Materials}
\newacronym{SBOM}{SBOM}{Software Bill of Materials}
\newacronym{EU}{EU}{European Union}
\newacronym{API}{API}{Application Programming Interface}
\newacronym{PoC}{PoC}{Proof-of-Concept}
\newacronym{PaaS}{PaaS}{Platform as a Service}
\newacronym{TEE}{TEE}{Trusted Execution Environment}
\newacronym{CSAM}{CSAM}{Child Sexual Abuse Material}
\newacronym{XAI}{XAI}{Explainable AI}
\newacronym{TAI}{TAI}{Trustworthy AI}
\newacronym{BOM}{BOM}{Bill of Materials}
\newacronym{ML}{ML}{Machine Learning}
\newacronym{TPM}{TPM}{Trusted Platform Module}
\newacronym{TDX}{TDX}{Trust Domain Extensions}
\newacronym{SGX}{SGX}{Software Guard Extensions}
\newacronym{PBAC}{PBAC}{Policy-Based Access Control}
\newacronym{RAM}{RAM}{Random-access Memory}
\newacronym{CPU}{CPU}{Central Processing Unit}
\newacronym{GPU}{GPU}{Graphics Processing Unit}
\newacronym{HSM}{HSM}{Hardware Security Module}
\newacronym{RCE}{RCE}{Remote Code Execution}
\AtBeginDocument{%
  }

\copyrightyear{2026}
\acmYear{2026}
\acmConference[CAIN '26]{5th International Conference on AI Engineering - Software Engineering for AI}{April 12--13, 2026}{Rio de Janeiro, Brazil}

\begin{document}

\title[AIBoMGen: Generating an AI Bill of Materials]{AIBoMGen: Generating an AI Bill of Materials for Secure, Transparent, and Compliant Model Training}

\author{Wiebe Vandendriessche}
\email{wiebe.vandendriessche@ugent.be}
\orcid{0009-0001-6499-0205}

%
%
%

\affiliation{%
    \institution{\textit{IDLab, Department of Information Technology}\\Ghent University - imec}
    \city{Ghent}
    \country{Belgium}
}

\author{Jordi Thijsman}
\orcid{0009-0007-2333-7051}
\affiliation{%
    \institution{\textit{IDLab, Department of Information Technology}\\Ghent University - imec}
    \city{Ghent}
    \country{Belgium}
}

\author{Laurens D'hooge}
\orcid{0000-0001-5086-6361}
\affiliation{%
    \institution{\textit{IDLab, Department of Information Technology}\\Ghent University - imec}
    \city{Ghent}
    \country{Belgium}
}

\author{Bruno Volckaert}
\orcid{0000-0003-0575-5894}
\affiliation{%
    \institution{\textit{IDLab, Department of Information Technology}\\Ghent University - imec}
    \city{Ghent}
    \country{Belgium}
}

\author{Merlijn Sebrechts}
\orcid{0000-0002-4093-7338}
\affiliation{%
    \institution{\textit{IDLab, Department of Information Technology}\\Ghent University - imec}
    \city{Ghent}
    \country{Belgium}
}

\renewcommand{\shortauthors}{Vandendriessche et al.}

\begin{abstract}
    The rapid adoption of complex \acrshort{AI} systems has outpaced the
    development of tools to ensure their transparency, security, and regulatory
    compliance. In this paper, the \acrfull{AIBOM}, an extension of the
    \acrfull{SBOM}, is introduced as a standardized, verifiable record of
    trained \acrshort{AI} models and their environments. A proof-of-concept
    platform, AIBoMGen, automates the generation of signed \acrshort{AIBOM}s by
    capturing datasets, model metadata, and environment details during
    training. The training platform acts as a neutral, third-party observer and
    root of trust. It enforces verifiable \acrshort{AIBOM} creation for every
    job. The system uses cryptographic hashing, digital signatures, and in-toto
    attestations to ensure integrity and protect against threats such as
    artifact tampering by dishonest model creators. Evaluations show that
    AIBoMGen reliably detects unauthorized modifications to all artifacts and
    can generate \acrshort{AIBOM}s with negligible performance overhead. These
    results highlight the potential of AIBoMGen as a foundational step toward
    building secure and transparent \acrshort{AI} ecosystems, enabling
    compliance with regulatory frameworks like the \acrshort{EU}'s
    \acrshort{AI} Act.
\end{abstract}


\begin{teaserfigure}
    \includegraphics[width=\textwidth]{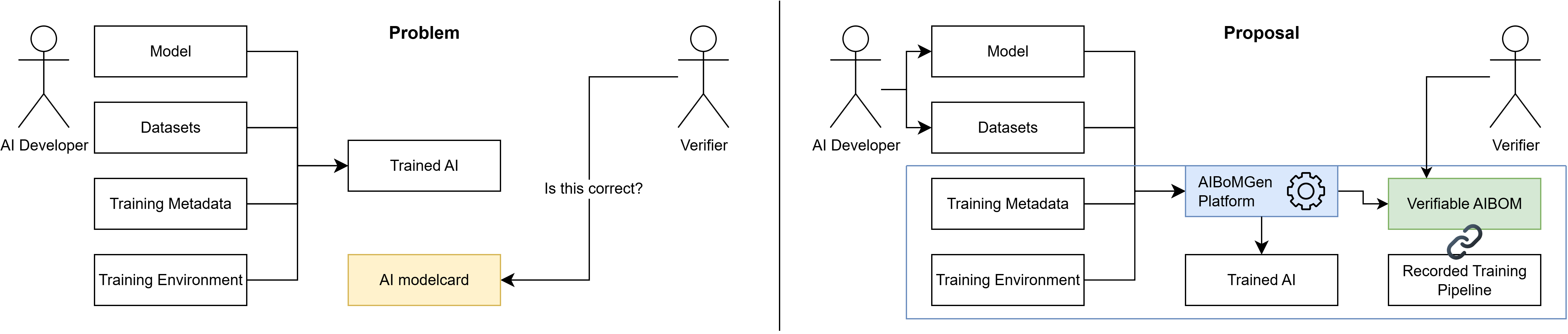}
    \caption{Overview of the proposed solution: AIBoMGen, a platform that generates verifiable AIBOMs while training the AI.}
    \Description{Overview of the proposed solution: Left: the problem with non-verifiable model cards. Right: AIBoMGen, a platform that generates verifiable AIBOMs while training the AI.}
    \label{fig:overview}
\end{teaserfigure}


\maketitle

\section{Introduction}

\acrshort{AI} models are black boxes, making it difficult for users to know essential
details such as where, how, and on what data a model was trained or tested.
These features are rarely visible from the model itself, yet are crucial for
determining whether an \acrshort{AI} model is safe, trustworthy, or appropriate for a
given scenario. The lack of transparency poses significant risks, especially
as \acrshort{AI} systems are increasingly deployed in sensitive and high-stakes
environments. Regulatory frameworks like the \acrshort{EU} \acrshort{AI} Act now require strict documentation
and accountability for \acrshort{AI} systems. This highlights the urgent need for
verifiable provenance and transparency tools \cite{eu_ai_2024}.

Transparency and explainability are now widely recognized as essential for
trust in \acrshort{AI} systems, with most organizations emphasizing these
principles in their ethical guidelines and linking explainability to
transparency to foster user trust \cite{balasubramaniam_transparency_2023}. The
global landscape of \acrshort{AI} ethics guidelines is expanding rapidly, with
over 80 documents issued in recent years by governments, corporations, NGOs,
and research groups worldwide \cite{jobin_global_2019}. This reflects a growing
concern about responsible \acrshort{AI} use. \acrfull{XAI} and \acrfull{TAI}
are crucial for building confidence in \acrshort{AI}. Provenance documentation
is emerging as a key solution, enabling the tracing of data, steps, and
contributors in \acrshort{AI} workflows to establish trust
\cite{kale_provenance_2023}. Despite these advances, many important models are
still getting trained on massive, under-documented datasets, leading to major
legal and ethical issues for developers \cite{longpre_position_2024}. For
example, Stanford University's investigation into open-source datasets revealed
many \acrfull{CSAM} images in the LAION-5B dataset, highlighting the risks of
insufficient oversight and documentation \cite{thiel_investigation_2023}. Model
cards \cite{ozoani_model_2022}, such as those used on Huggingface
\cite{hugging_face_hugging_2025}, are intended to improve transparency by
documenting model functionality and usage, but since they are manually written
by creators, their completeness and accuracy cannot be guaranteed. Without
cryptographic verification or external oversight, model cards remain
self-declared statements rather than verifiable evidence.

Similar transparency breakdowns are visible in commercial AI deployments.
Opaque training data practices in large generative models have triggered a
growing list of legal disputes and uncertainty around copyright and
intellectual property. Downstream users struggle to assess whether tools can be
used safely and lawfully \cite{marr_examples_2024}. In high-stakes domains such
as finance and healthcare, non-transparent models have led to unexplained
credit denials or misdiagnoses, and organizations report difficulty assembling
the documentation now expected by regulators and customers
\cite{marr_examples_2024}. Taken together, these examples illustrate typical
challenges faced by organizations in practice: many practitioners lack
practical mechanisms to produce standardized, verifiable evidence about how
models were trained and configured in real-world systems.

The \acrfull{AIBOM}, an extension of the \acrfull{SBOM}, has emerged as part of
the solution. \acrshort{AIBOM}s provide a structured inventory of all
components involved in \acrshort{AI} model development \cite{ghosh_ai_2024}.
However, transparency alone is not enough; trust is needed
\cite{techworks_trustworthy_2023}. Without mechanisms to ensure the integrity
and authenticity of the documented information, \acrshort{AIBOM}s cannot
effectively support compliance and security \cite{techworks_compliance_2024}.
Many \acrshort{AI} development environments and research-focused
infrastructures do not have a standardized procedure in place to provide
guarantees on these \acrshort{AIBOM}s, let alone generate them
\cite{imec_gpulab_2025}.

\textbf{AIBoMGen} \cite{vandendriessche_aibomgen_2025} (Fig. \ref{fig:overview}) is introduced as
an open-source\footnote{https://github.com/idlab-discover/AIBoMGen} training
platform that automates the generation of signed \acrshort{AIBOM}s. Instead of relying on
the model provider or their training code to honestly report properties of the
model, our platform functions as a neutral, third party observer and root of
trust that captures metadata while the training runs: input files (datasets,
base model), configuration (epochs, batch size, framework), environment snapshot
(container image, resource metrics) and outputs (model, metrics, logs). It enforces
verifiable \acrshort{AIBOM} creation for every job.

All artifacts are hashed and linked through an in-toto based attestation.
In-toto is a framework for supply chain security that generates
cryptographically signed records of each step in a process, enabling end-to-end
verification of integrity and authenticity \cite{in-toto_-toto_2025-1}. The
resulting \acrshort{AIBOM} and its signature can be verified via \acrshort{API}
endpoints by any party using the platform public key. It is important to note
that in this case the \acrshort{AIBOM} only covers the training step and
represents the state of a trained model and its environment. The loosely
coupled architecture is built on in-toto principles and can be extended to
include the deployment step in future work as well.

In short, AIBoMGen turns \acrshort{AI} training metadata from a manual,
error‑prone exercise into a built-in platform feature. This reduces the gap
between current informal model documentation and the stronger evidence needed
for audit, risk review and regulatory reporting.

\subsection{Research Questions}
\begin{itemize}
    \item \textbf{RQ1:} How can trust be effectively integrated into the \acrshort{AI}
          lifecycle to ensure transparency, integrity, and ethical concerns?

    \item \textbf{RQ2:} To what extent can trust mechanisms be integrated
          without excessively limiting developer flexibility and innovation?

    \item \textbf{RQ3:} Which concrete methods and technologies enable the
          automated generation of \acrshort{AIBOM}s while maintaining their
          trustworthiness?
\end{itemize}

\subsection{Contributions}
\begin{itemize}
    \item Presents AIBoMGen, a novel platform that automates the generation of
          cryptographically verifiable \acrshort{AI} Bills of Materials
          (\acrshort{AIBOM}s) during \acrshort{AI} model training.

    \item Demonstrates a practical approach to verifiable \acrshort{AI} models
          using in-toto attestations, artifact hashing, and digital signatures
          without the use of \acrshort{TEE}s.

    \item Evaluates AIBoMGen's performance, storage impact, and security
          through various experiments and threat analysis.

    \item Provides a proof-of-concept implementation and open-\\source
          reference for automated, transparent, and verifiable \acrshort{AI}
          model documentation.
\end{itemize}

\section{Related Work}

Different frameworks and standards have been developed to address provenance,
integrity, and compliance in \acrshort{AI} workflows. This section reviews the
current \acrshort{AIBOM} standards and related frameworks.

\subsection{Current State of AIBOM}

\acrfull{SBOM} standards like CycloneDX \cite{owasp_foundation_cyclonedx_2024}
and SPDX 3.0 \cite{linux_foundation_spdx_2023} are beginning to adapt their
\acrshort{SBOM} schemas for \acrshort{AI}-specific components and workflows.
CycloneDX has been extended to include \acrshort{AI}-specific components,
enabling structured and machine-readable \acrshort{AIBOM}s
\cite{owasp_foundation_cyclonedx_2024-1}. Similarly, SPDX 3.0 introduces
fields for \acrshort{AI}-specific metadata \cite{linux_foundation_spdx_2024}
and compliance with regulations like the \acrshort{EU} \acrshort{AI} Act
\cite{linux_foundation_implementing_2024}. These standards provide a foundation
for creating interoperable and verifiable \acrshort{AIBOM}s, but their
adoption remains limited.

Despite their importance, \acrfull{SBOM}s and \acrshort{AIBOM}s are not widely
adopted due to limited awareness and a lack of mature, easy-to-use tools
\cite{stalnaker_boms_2024}. Also, the manual effort required to maintain them
is a significant barrier. According to Stalnaker et al., only about half of the
respondents are familiar with \acrshort{SBOM}s~\cite{stalnaker_boms_2024}. The
existence of multiple, sometimes incompatible, standards like SPDX and
CycloneDX can also confuse developers and cause interoperability issues. Once
created, \acrshort{SBOM}s must be maintained as software changes, which is a
common challenge. There is a clear need for tools that can automatically update
\acrshort{SBOM}s and \acrshort{AIBOM}s as changes occur. Integrating
\acrshort{SBOM} and \acrshort{AIBOM} creation directly into existing systems,
such as build tools and package managers, and developing better,
language-specific tools would help increase adoption. Some suggest that
mandates could help clarify the benefits and costs of using these tools
\cite{stalnaker_boms_2024}.

Most mainstream \acrshort{AI} training platforms prioritize usability and
flexibility over verifiability and auditability. For instance, GPULab offers
researchers GPU access through containerized workloads on a distributed testbed
infrastructure \cite{tomer_experimenting_2022}. While this enables flexible
deployment of Docker-based \acrshort{ML} experiments, it provides no mechanisms
to verify which datasets, code, or configurations were used, nor does it
capture the training lineage or protect against post hoc modifications.

This absence of verifiability is not unique to GPULab. Commercial platforms
such as Google Colab, AWS SageMaker, and Kaggle Notebooks similarly allow for
scalable model training but do not include native support for provenance
tracking or supply chain integrity. In these systems, all trust is implicitly
placed in the developer, with no cryptographic evidence or transparency
artifacts generated during training. As a result, any downstream claims about
model performance, fairness, or regulatory compliance remain unverifiable by
external parties.

\subsection{Related Frameworks and Initiatives}

At the time of writing, many \acrshort{AIBOM}-related projects have emerged to
address the challenges of documenting, verifying, and attesting to the
properties of AI.

\paragraph{Manifest Cyber} One proposal on \acrshort{AIBOM} content, from Manifest Cyber
\cite{bardenstein_whitepaper_2023}, states that the following should be
included:
\begin{itemize}
    \item Metadata of the model, such as: name, version, description and
          authorship.

    \item Timestamp and generation tools.

    \item Detailed information about model architecture, datasets used and
          ethical considerations and intended usage.

    \item Dependencies required by the model such as: external\\ sources and
          \acrshort{ML} libraries.
\end{itemize}

While Manifest Cyber's example is a strong starting point, it has several key
limitations. The \acrshort{BOM} relies on declarative documentation similar to
Huggingface model cards, which can be manually edited and is not
cryptographically verifiable. There are no artifact hashes or digital
signatures, making it impossible to verify the authenticity or integrity of the
documented files. Key information, such as intended use, misuse scenarios, and
limitations, relies solely on the developer’s honesty and accuracy. This
increases the risk of inconsistencies between the documented information and
the actual properties of the model.

\paragraph{ALOHA}
The ALOHA framework generates post hoc \acrshort{AIBOM}s for Hugging Face
models by parsing metadata and dependencies associated with uploaded artifacts
\cite{davino_aloha_2025}. However, the integrity of these outputs is limited by
the assumption that the uploaded model accurately reflects the true training
process. Since the system does not instrument or observe training directly, it
cannot detect tampering, substitution, or unreported preprocessing.

\paragraph{Atlas}
Recent research has also explored the use of \acrfull{TEE}s to create secure
and verifiable \acrshort{AIBOM}-like metadata. For example, the Atlas project
captures \acrshort{AI} lifecycle provenance using Intel \acrshort{TDX}. It logs
training and inference steps into an immutable transparency ledger and produces
attestations for model consumers \cite{spoczynski_atlas_2025}. While this
approach provides strong integrity guarantees under the assumption of a trusted
enclave, reliance on specialized hardware limits portability. The framework
does not use standardized \acrshort{AIBOM} schemas and lacks support for easy
integration into existing \acrshort{ML} pipelines.

\paragraph{Laminator}
This framework generates verifiable \acrshort{ML} property cards using
\acrshort{TEE}-based instrumentation, enabling attestations about model
training, inference behavior, and statistical properties
\cite{duddu_laminator_2025}. This approach enforces confidentiality and
integrity but does not expose training internals in a transparent or auditable
way. The resulting cards are tied to a proprietary format and cannot be readily
interpreted outside the Laminator ecosystem. Moreover, the framework is not
structured as a cloud-native or service-oriented platform and cannot be
trivially adopted in production pipelines.

\paragraph{AICert}
The AICert framework uses \acrshort{TPM} boot measurements to bind training
configuration, dataset hashes, and output weights into a hardware-rooted proof
\cite{mithril_security_aicert_2025}. It supports fine-tuning within its
predefined Axolotl container. It provides strong platform integrity but only
coarse runtime visibility and outputs a custom (\acrshort{AIBOM}-like) proof
file.

Across these systems, key limitations persist. First, most frameworks rely on
\acrshort{TEE}-based security, which introduces hardware dependencies, cost,
and platform lock-in, while assuming near-perfect enclave security despite a
history of side-channel vulnerabilities in \acrshort{SGX} and \acrshort{TDX}.
Second, the majority do not emit artifacts in standardized machine-readable
formats (e.g., CycloneDX / SPDX \acrshort{AIBOM}), limiting interoperability.
Finally, few expose programmatic \acrshort{API}s for external verification or
pipeline integration, reducing usability outside research contexts.

\section{AIBoMGen System Design}

To further the state of the art in trustable statements about \acrshort{AI}
model supply chains, AIBoMGen is proposed. This platform ensures trustable
claims in \acrshort{AIBOM}s by introducing the following key features:

\begin{enumerate}
    \item \textbf{Verifiable claims:} The \acrshort{PoC} introduces in-toto attestations
          \cite{in-toto_-toto_2025-1}, providing cryptographic evidence of the
          training process and environment.

    \item \textbf{Artifact hashes for integrity:} These attestations include
          SHA-256 hashes of artifacts such as datasets, models, and metrics, enabling
          users to verify the integrity of each item.

    \item \textbf{Digital signatures:} The \acrshort{AIBOM} itself is signed,
          and signature fields are embedded in the \acrshort{BOM}. This allows one
          to prove that training was done on the platform and that the
          \acrshort{AIBOM} was not tampered with after generation.

    \item \textbf{Automation:} The process is automated, reducing the risk of
          human error or manipulation.
\end{enumerate}

In contrast to related projects, our approach uses the platform (a locked-down
\acrshort{API}) as the root of trust instead of not trusting the platform and
requiring immediate \acrshort{TEE} use. Users can only upload inputs and
parameters, making sure there is no remote shell or code execution
(\acrshort{RCE}). A fixed workflow in managed worker containers performs
training, mandatory metadata capture, hashing and signing, then outputs a
signed \acrshort{AIBOM} with in-toto link, describing the artifacts and supply
chain. This blocks remote arbitrary code, lowers early overhead (no enclave
provisioning or quote handling) and preserves an upgrade path: later a
\acrshort{TEE} can wrap the worker to add hardware-backed environment/property
attestations using the same schema. The trade‑off of AIBoMGen is reduced
developer freedom.

Allowing \acrshort{RCE} on worker nodes would fundamentally undermine the trust
guarantees of the platform. If users or attackers could execute arbitrary code
during training, they could bypass attestation, falsify the \acrshort{AIBOM},
or tamper with artifacts, making it impossible to guarantee the integrity of
the training process.

A detailed threat evaluation of the platform's security and trust guarantees is
provided in Section~\ref{subsec:threat_evaluation}.

Custom scripts or novel machine learning frameworks are simply not possible on
the platform yet. Only pre-approved training workflows, currently limited to
TensorFlow-based classification and regression tasks, are available in the
controlled pipeline. To support additional frameworks or custom scripts, these
must first be explicitly integrated and vetted as part of the platform’s
managed workflow. This restriction ensures that all training steps are
observable and verifiable. In future iterations, adding \acrshort{TEE}s could
allow for securely running a broader range of user code inside attested
environments, shifting trust from strict pipeline enforcement to
hardware-backed isolation.

The platform acts as a scalable \acrshort{AI} \acrfull{PaaS} and could be
extended and deployed on cloud infrastructure, backed by a governmental
consumer protection agency. However, the current system should be considered a
\acrshort{PoC} and is not intended for production use.

This \acrshort{PoC} should therefore be read primarily as a reference
architecture for integrating observation, attestation, signing, and
verification into training pipelines, rather than as a production-ready
service. The deliberately strict, locked-down workflow isolates the trust and
provenance mechanisms in a controlled setting, clarifying the trade-off between
strong guarantees and developer flexibility. During design, more permissive
instrumentation of existing notebook or MLOps environments and
\acrshort{TEE}-first architectures were considered, but these alternatives
would either allow unobserved steps or tie the approach too tightly to specific
hardware assumptions.

\subsection{Architecture}
\label{subsec:architecture}

\begin{figure}[tbp]
    \centerline{ \includegraphics[width=\columnwidth]{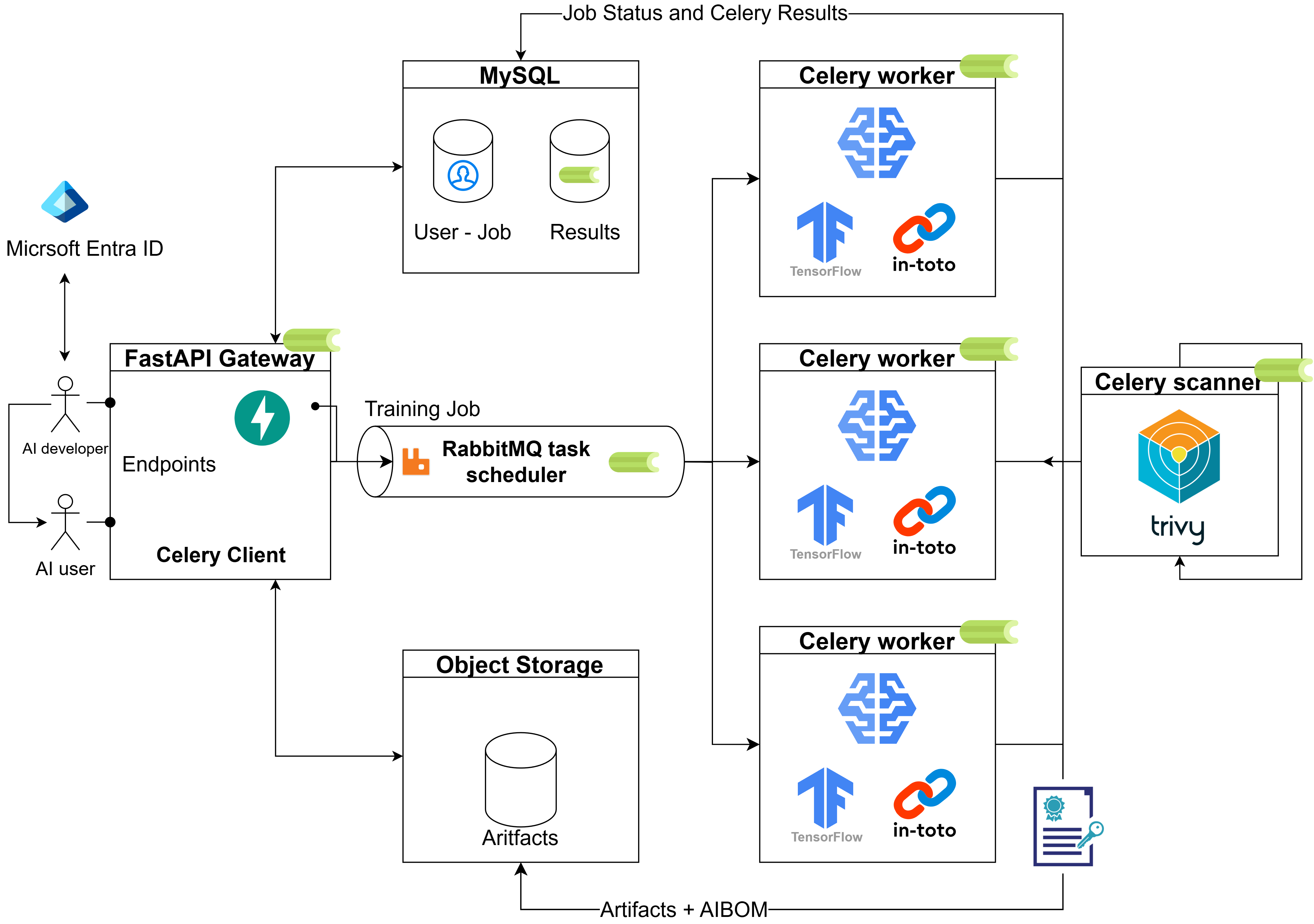}}
    \caption{Architecture diagram of AIBoMGen backend, authorized entities can
        submit training jobs which are executed and attested by the platform.}
    \Description{The diagram shows the backend architecture of AIBoMGen, including the API gateway, task scheduler, worker nodes, and data storage. Arrows indicate the flow of job submissions, artifact generation, and attestation processes. The figure clarifies how the platform enforces transparency and trust in the AI training pipeline.}
    \label{fig:architecture}
\end{figure}

The AIBoMGen platform is designed as a modular, scalable, and secure system
that automates the generation of cryptographically verifiable \acrshort{AI}
Bills of Materials (\acrshort{AIBOM}s). Its primary objective is to enforce
transparency and trust in the \acrshort{AI} training pipeline through an
attested, auditable infrastructure. Fig.~\ref{fig:architecture} displays a
simplified architecture diagram. The system consists of several key components,
each playing a specific role in the \acrshort{AIBOM} generation and
verification process:

\paragraph{\acrshort{API} Gateway} The \acrshort{API} Gateway serves as the central point of interaction
for both developers and verifiers. It handles job submissions, artifact
retrieval, and verification requests. By limiting interaction only through the
gateway, remote code execution on the workers becomes impossible, without a
specific vulnerability elsewhere. The implementation used the FastAPI framework
\cite{fastapi_fastapi_2025}, which automatically provides an OpenAPI
specification and interactive documentation.

\paragraph{Task Scheduler} A central task scheduler, implemented using Celery \cite{solem_celery_2023},
coordinates the execution of training jobs and other tasks across the system.
It uses RabbitMQ \cite{rabbitmq_rabbitmq_2025} as a message broker to ensure
reliable communication. MySQL is used as a results backend
\cite{oracle_mysql_2025}.

\paragraph{Worker Nodes} Worker nodes are responsible for executing training jobs in isolated Docker
containers. Each worker processes tasks from the task scheduler and generates
the \acrshort{AIBOM} after training. They also perform environment extraction,
artifact hashing, and in-toto generation to ensure the integrity of the
training process. The scheduler combined with the worker nodes are meant to
simulate a production distributed \acrshort{AI} training system, where multiple
workers can process jobs concurrently.

\begin{figure}[tbp]
    \centerline{\includegraphics[width=0.7\columnwidth]{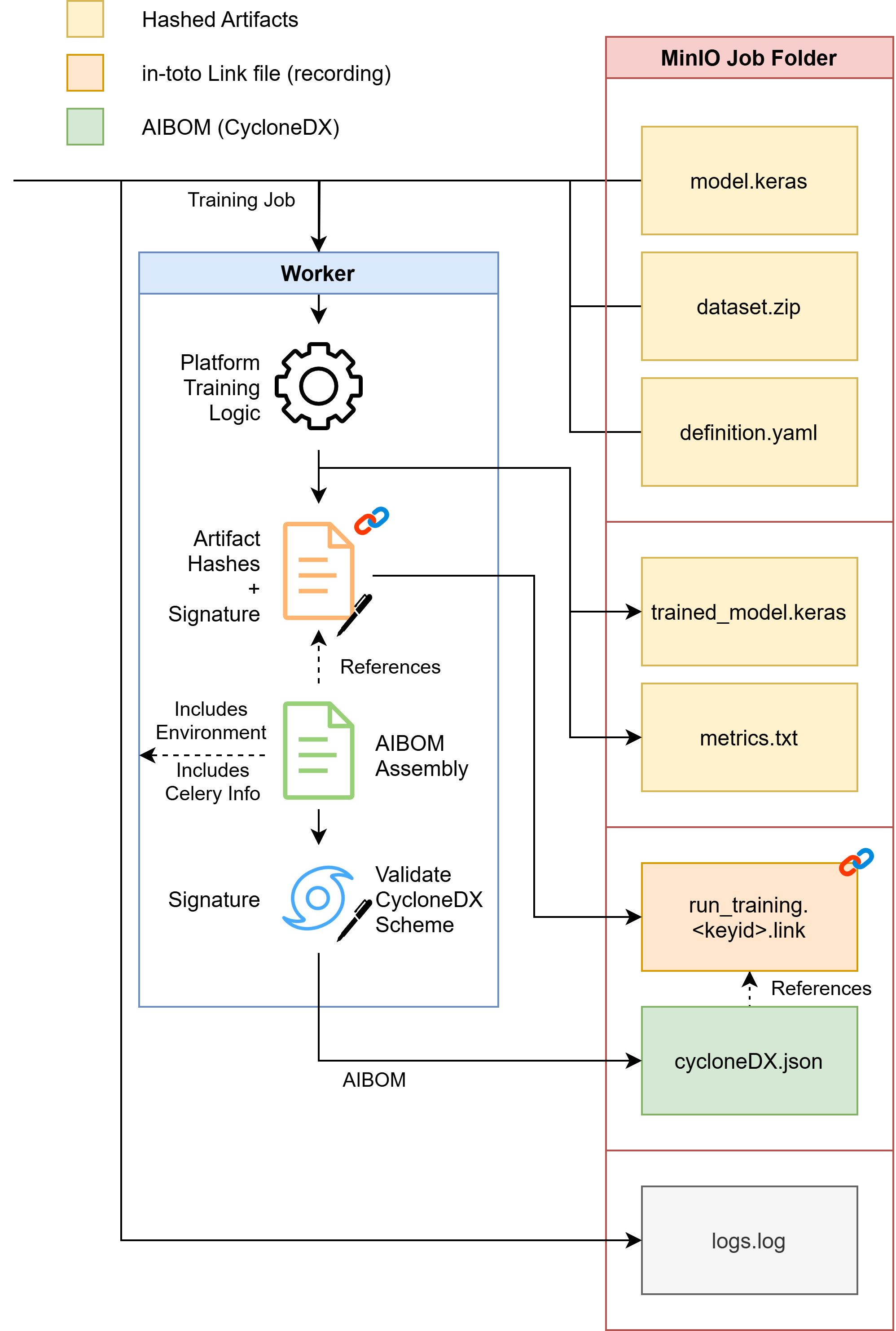}}
    \caption{Overview of the main steps performed by a worker node during a
        training job.}
    \Description{This figure details the sequence of actions taken by a worker node, including downloading inputs, running training, capturing metadata, hashing artifacts, generating and signing attestations, and uploading results. It visually represents the attestation and verification process for each training job.}
    \label{fig:worker_intoto}
\end{figure}

Figure~\ref{fig:worker_intoto} illustrates the main steps executed by a worker
node. After receiving a job from the scheduler, the worker downloads the
required input files (datasets, model, configuration). It then runs the
training process inside the container, capturing relevant metadata and
environment details. Once training is complete, the worker hashes all
artifacts, includes them in a signed in-toto link file, generates the
\acrshort{AIBOM} and signs the \acrshort{AIBOM}. Finally, all results are
uploaded to the storage, and the worker reports completion to the scheduler.
This workflow ensures that every step is recorded and verifiable.

\paragraph{Data Storage} MinIO, an S3-compatible object storage solution \cite{minio_minio_2025}, is
used to store datasets, models, logs and generated \acrshort{AIBOM}s. It
ensures secure and scalable storage for all artifacts. The system uses
presigned URLs to provide time-limited access to stored artifacts. The storage
is deemed secure because only the platform has write access to the MinIO
instance, and users can only read their own files through presigned URLs. In
addition to that, MinIO has an option to enable \acrfull{PBAC}, which can be
used to further restrict access to certain files.

\paragraph{Vulnerability Scanner} The platform is designed to allow the integration of different compliance
auditing and certification tools as needed. In the current implementation, a
dedicated Trivy scanner \cite{aqua_security_trivy_2025} periodically checks the
Docker images used by the workers for vulnerabilities, with results stored in
the object store and summarized in the \acrshort{AIBOM}. However, other tools
can easily be added or substituted. For example, a compliance checker could
periodically scan all finished jobs in MinIO, analyze the generated
\acrshort{AIBOM}s and vulnerability reports, and automatically determine
whether each job meets requirements such as the \acrshort{AI} Act or internal
policies. This enables the platform to support both the training and
certification steps of the \acrshort{AI} lifecycle.

\subsection{User flows}
\begin{figure}[tbp]
    \centerline{\includegraphics[width=\columnwidth]{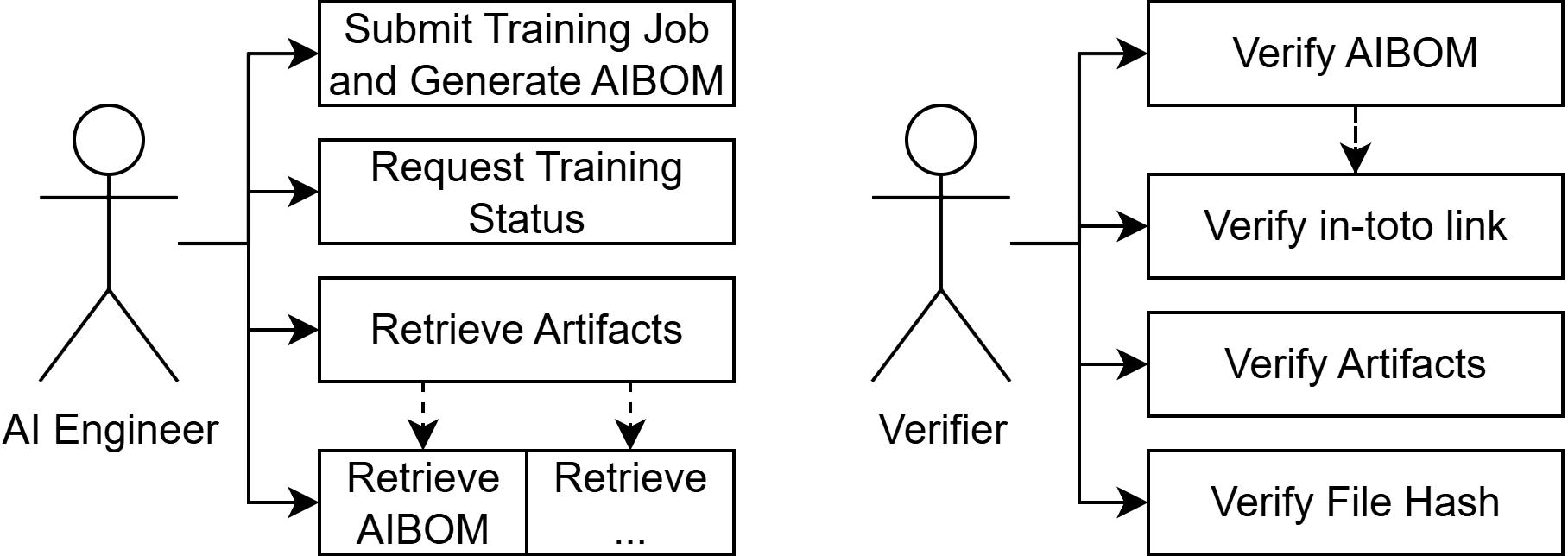}}
    \caption{Use case diagram of the AIBoMGen application.}
    \Description{The use case diagram depicts the interactions between the two main actors: the AI Engineer and the Verifier. It shows developers submit jobs and retrieve artifacts, while verifiers check the integrity and authenticity of outputs using the platform’s verification endpoints.}
    \label{fig:usecase}
\end{figure}

The system involves two primary actors: the \acrshort{AI} Engineer and the
Verifier. Fig.~\ref{fig:usecase} illustrates the main system use cases of these
actors.

The \acrshort{AI} Engineer submits training jobs to the system through a secure
\acrshort{API}. These jobs include the model, dataset, and training parameters.
Currently, the implementation supports classification and regression tasks on
tabular and image data using TensorFlow. Once the training is complete, the
developer retrieves resulting artifacts, including a trained model, metrics,
and the generated \acrshort{AIBOM}. The developer can make the \acrshort{AIBOM}
publicly available for transparency.

The verifier ensures the integrity and authenticity of the \acrshort{AIBOM} and
its associated artifacts. This is achieved by verifying cryptographic
signatures and hashes recorded during the training process. The verifier can
also validate artifact hashes against the hashes stored in the \acrshort{AIBOM}
to detect any tampering.

The \acrshort{PoC} implementation has four verification endpoints:
\begin{itemize}
    \item \textbf{Verify \acrshort{AIBOM} and in-toto Link File:} The
          verifier uploads the signed CycloneDX \acrshort{AIBOM} to the \acrshort{API}
          Gateway. The endpoint validates the \acrshort{BOM}'s schema, verifies
          its cryptographic signature, and checks the referenced attestation
          file (in-toto link). This provides an end-to-end integrity and authenticity
          check for the entire training process and its outputs.

    \item \textbf{Verify in-toto Link File:} The verifier uploads the
          attestation file to the \acrshort{API} endpoint, which checks the
          cryptographic signatures, layout validity, and artifact rules. This ensures
          the file is authentic and the recorded actions were performed as expected.
          The in-toto documentation provides more details about how in-toto can be used for supply chain security \cite{in-toto_-toto_2025}. This endpoint is useful when only the attestation file is available or when
          CycloneDX \acrshort{AIBOM} validation is performed separately.

    \item \textbf{Verify File Hash Against Link File:} The verifier uploads
          both the in-toto link file and a specific artifact (e.g., dataset,
          model, metrics ...) to the \acrshort{API}. The endpoint computes the
          hash of the artifact and compares it to the hash recorded in the
          link file, confirming the artifact is the same as the one recorded.

    \item \textbf{Verify All Artifacts in Storage:} The verifier uploads the
          in-toto link file to the \acrshort{API}, which then automatically
          downloads all referenced materials and products from storage. The endpoint
          computes their hashes and compares them to those in the link file,
          detecting any tampering in storage.
\end{itemize}

The verifier uses the results from the "Verify \acrshort{AIBOM}" endpoint to
make a trust decision. Optionally, they can further validate artifacts or
certain artifact hashes and vulnerabilities to ensure the integrity of the
training process.

In the current \acrshort{PoC}, all attested artifacts are signed with a single
platform-managed key pair, and its public key is shared with verifiers. In a
more production-like deployment, this can evolve to tenant- or project-specific
keys, where each approved organization controls its own private signing key
(e.g., in an \acrshort{HSM}) and registers the corresponding public key with
the platform. The same approach can be applied across the full lifecycle so
that later stages, such as deployment, are also attested and signed,
potentially by distinct keys on external deployment platforms.

\section{Evaluation and Results}
AIBoMGen was evaluated on three key metrics: performance overhead, storage
footprint, and security. The results demonstrate that the platform introduces
minimal performance and storage overhead while providing robust security
measures. This section provides an analysis of these metrics, along with
additional insights into the threat evaluation.

\subsection{Test Environment}
All experiments were conducted with the backend running on a host machine
equipped with an AMD Ryzen 7 5800H \acrshort{CPU}, 32GB \acrshort{RAM}, and a
NVIDIA GeForce RTX 3060 \acrshort{GPU}. Most tests were performed by submitting
training jobs manually and via a series of Jupyter Notebooks, which are
published together with results on
Zenodo~\cite{vandendriessche_aibomgen-experiments_2025}. The latest version of
the experiments is available on
GitHub\footnote{\url{https://github.com/idlab-discover/AIBoMGen-experiments}},
including code and instructions for replication.

\subsection{Performance Evaluation}

\begin{figure}[tbp]
    \centerline{\includegraphics[width=\columnwidth]{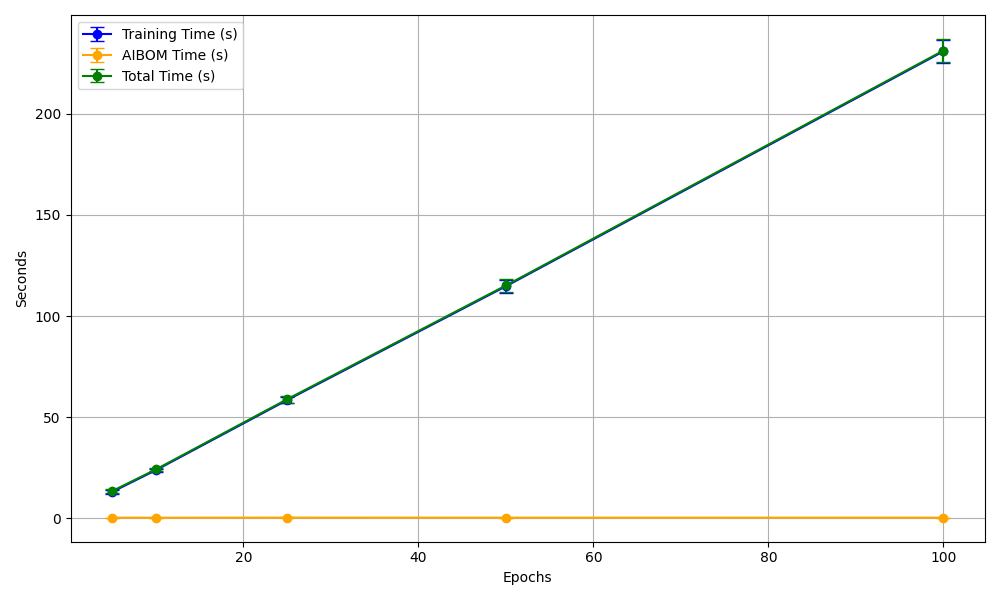}}
    \caption{Total training time and AIBOM generation time as a function of epoch count.}
    \Description{This line chart plots the total training time and AIBOM generation against the number of epochs, demonstrating AIBOM generation time is not dependent on training. The underlying data is available in Table~\ref{tab:aibom_performance}.}
    \label{fig:epoch_performance}
\end{figure}

\begin{figure}[tbp]
    \centerline{\includegraphics[width=\columnwidth]{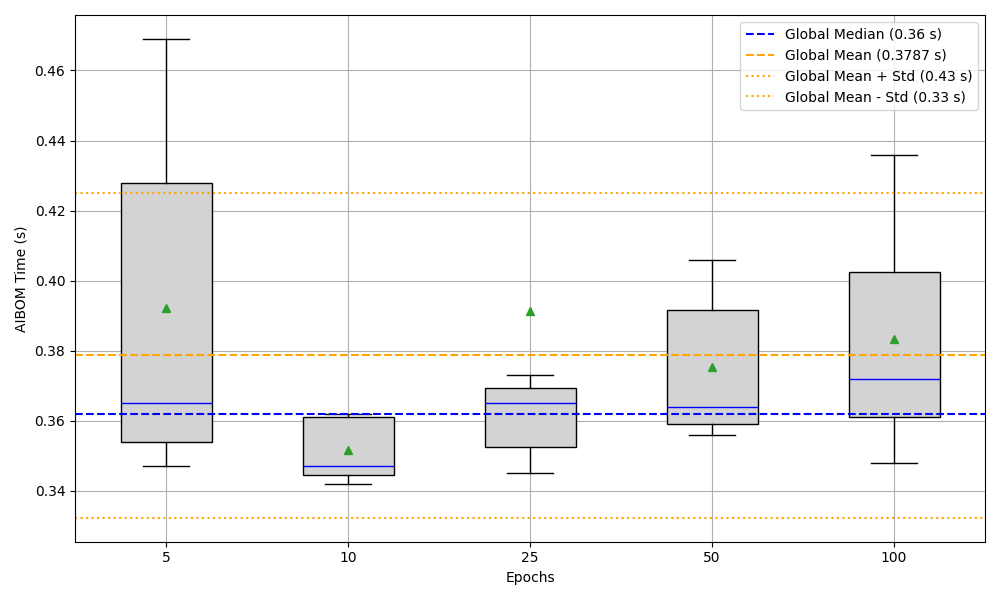}}
    \caption{AIBOM generation time box plots across different epoch counts
        with global mean and std.}
    \Description{Box plots show the distribution of AIBOM generation times for various epoch counts, with overlays for global mean and standard deviation. The underlying data is available in Table~\ref{tab:aibom_performance}.}
    \label{fig:aibom_time}
\end{figure}

The performance impact of \acrshort{AIBOM} generation was evaluated by running
training jobs with varying epochs. As shown in Fig.~\ref{fig:epoch_performance}
and Table \ref{tab:aibom_performance}, training time increases linearly with
the number of epochs, while \acrshort{AIBOM} generation time remains nearly
constant (Fig.~\ref{fig:aibom_time}). The generation step is fast, with only
minor fluctuations from processor variability. This negligible overhead is
expected, since AIBoMGen decouples metadata capture and attestation from
training. These results show that \acrshort{AIBOM} generation is practical for
real-world training workflows without affecting performance.


\begin{table*}[tbp]
    \caption{Training and \acrshort{AIBOM} Generation Times (mean and std) for Different Epochs.}
    \label{tab:aibom_performance}
    \centering
    \begin{tabular}{lcccccc}
        \toprule
        \textbf{Epochs} & \textbf{Training Time (mean s)} & \textbf{(std s)} & \textbf{AIBOM Time (mean s)} & \textbf{(std s)} & \textbf{Total Time (mean s)} & \textbf{(std s)} \\
        \midrule
        5               & 13.110                          & 1.110            & 0.392                        & 0.049            & 13.503                       & 1.156            \\
        10              & 23.865                          & 0.704            & 0.352                        & 0.009            & 24.216                       & 0.712            \\
        25              & 58.502                          & 1.708            & 0.391                        & 0.086            & 58.893                       & 1.757            \\
        50              & 114.735                         & 3.167            & 0.375                        & 0.021            & 115.111                      & 3.182            \\
        100             & 230.817                         & 5.748            & 0.383                        & 0.032            & 231.200                      & 5.751            \\
        \bottomrule
    \end{tabular}
\end{table*}

\subsection{Storage Impact}
\begin{figure}[tbp]
    \centerline{ \includegraphics[width=\columnwidth]{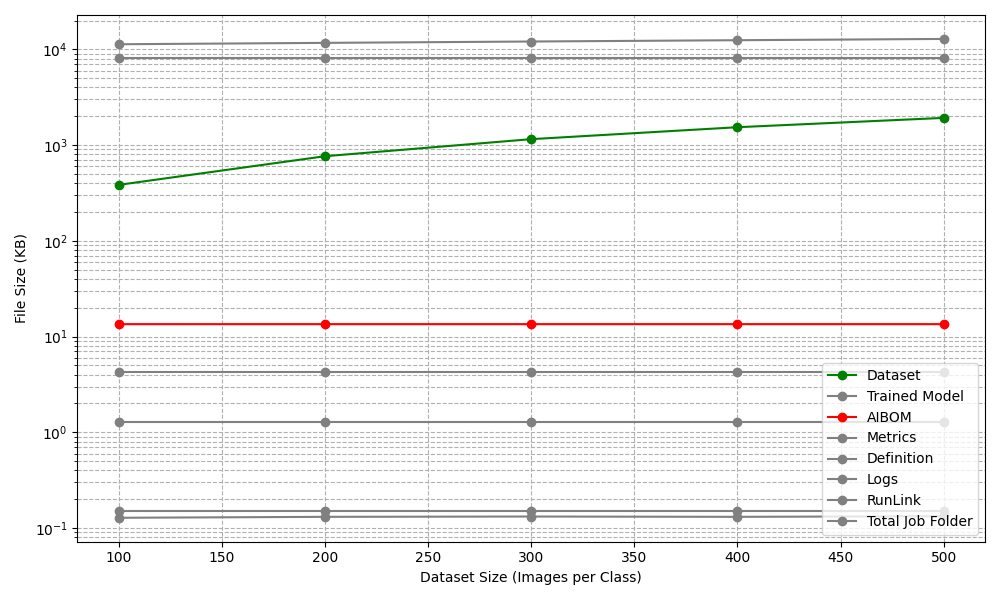}}
    \caption{Plot showing artifact file sizes for different dataset sizes.}
    \Description{This figure compares the file sizes of datasets, models, and AIBOMs for different dataset sizes, illustrating that AIBOMs remain small relative to other artifacts and are almost independent to dataset sizes.}
    \label{fig:storage_dataset}
\end{figure}

\begin{figure}[tbp]
    \centerline{ \includegraphics[width=\columnwidth]{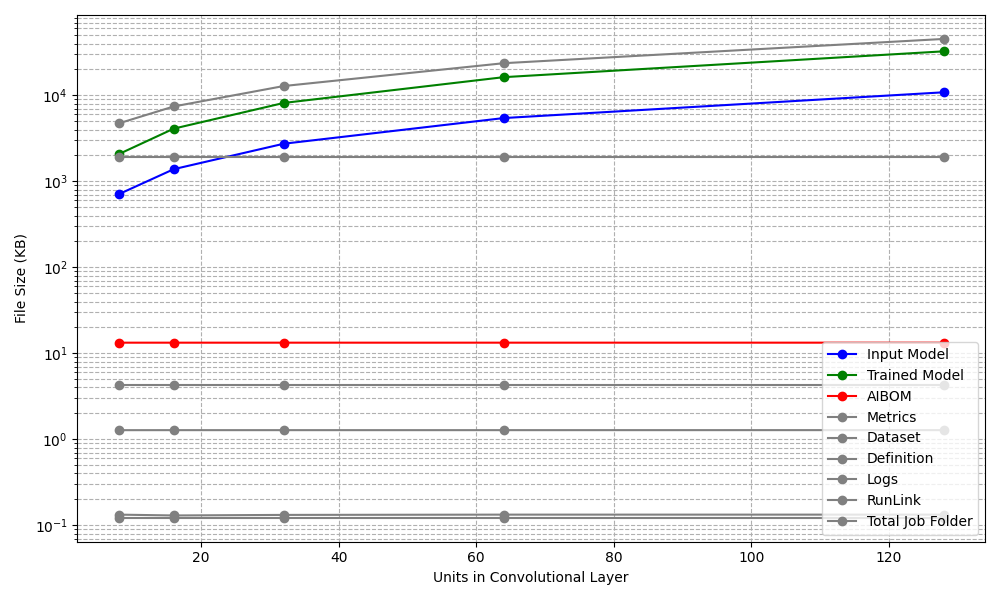}}
    \caption{Plot showing artifact file sizes for different input model
        sizes.}
    \Description{This figure compares the file sizes of datasets, models, and AIBOMs for different model sizes, showing that AIBOMs add minimal storage overhead compared to model files and are almost independent to model files.}
    \label{fig:storage_model}
\end{figure}

The storage overhead of \acrshort{AIBOM} generation was evaluated by analyzing
the sizes of artifacts produced during training. Fig.~\ref{fig:storage_dataset}
shows that the \acrshort{AIBOM} size grows slightly with dataset size but
remains small relative to the dataset and model sizes. Similarly,
Fig.~\ref{fig:storage_model} demonstrates that the \acrshort{AIBOM} size is
minimal compared to the trained model size, even for larger models. These
results confirm that the storage impact is negligible when compared to the
overall artifacts involved, especially for very large jobs.

\subsection{Threat Evaluation}
\label{subsec:threat_evaluation}

The threat evaluation of AIBoMGen is structured around several threat classes,
each grouping related attack scenarios and their countermeasures. This approach
clarifies how the platform addresses different categories of risk and where
vulnerabilities remain. These classes broadly follow common structures used in
software supply chain security discussions (inputs, process, artifacts, access,
and underlying platform), but are tailored to the specifics of AIBoMGen.

\paragraph{Input Tampering}
The first class covers attacks that attempt to compromise the system by
manipulating user-supplied files. For example, a malicious actor could embed a
Lambda layer in a Keras model file, enabling arbitrary code execution when the
model is loaded \cite{righi_tensorflow_2022}. To counter this, AIBoMGen
enforces safe model loading for Keras models
\cite{tensorflow_tfkerasmodelsload_model_2024}, restricts model loading to
isolated Docker containers, and includes model hashes in the \acrshort{AIBOM}.
However, if TensorFlow's safe model is bypassed or a model contains
undetectable malicious logic, these measures may not suffice. It is important
to verify whether the current TensorFlow API still allows custom layers or
arbitrary code in serialized models. If so, adopting safer formats such as
Safetensors, which is used by Huggingface to prevent arbitrary code execution,
would further mitigate this risk. Similarly, tampering with datasets or
definition files, such as injecting malformed data, can disrupt training
results. The platform mitigates this by validating dataset structure and
content, securely extracting archives, and processing all data in isolated
containers. Still, malicious but valid datasets (e.g., poisoned or biased data)
will evade detection, as the system does not check for data quality or intent.

To further reduce risks from malicious datasets or model files, future versions
could integrate other safe serialization formats and advanced dataset
validation tools, including external auditing or AI-based mechanisms to detect
poisoned or biased data.

\paragraph{Process Tampering}
This class includes threats that target the integrity of the training workflow
itself. An attacker might attempt to modify training parameters, inject code,
or alter the environment to corrupt the process. AIBoMGen addresses this by
running jobs in non-root, read-only containers with minimal privileges, and by
using in-toto attestation and CycloneDX \acrshort{BOM}s to record the
environment and all steps. Nevertheless, if a container escape or privilege
escalation occurs, the process could still be compromised. Another risk is the
bypassing of \acrshort{AIBOM} generation, which would allow unrecorded or
unverifiable models. The system enforces \acrshort{AIBOM} creation for every
job and restricts all interactions to a secure \acrshort{API}, but
vulnerabilities in the \acrshort{API} or worker logic could still allow bypass
attempts.

Stronger isolation layers, such as using more restrictive container runtimes
(e.g., gVisor, Kata Containers) or runtime monitoring solutions, could help
detect and prevent container escapes or privilege escalations. Regular code
integrity verification, by including hashes of worker and API code in the
\acrshort{AIBOM}, could further protect against unauthorized modifications.

\paragraph{Artifact Tampering}
Once artifacts such as models, logs, metrics, or \acrshort{AIBOM} files are
generated, an attacker could try to modify them before they are distributed.
AIBoMGen mitigates this by hashing all artifacts, storing them in
access-controlled MinIO buckets, and cryptographically signing both the
\acrshort{AIBOM} and in-toto files. Users can verify the integrity and
authenticity of all outputs. While cryptographic signatures and hash
verification protect against falsified \acrshort{AIBOM}s or fake link files,
these safeguards could be undermined if the private signing keys were ever
compromised.

The risk of private signing key compromise can be mitigated by storing keys in
\acrfull{HSM}s and by implementing regular key rotation and secure key
management policies.

\paragraph{Access and Authenticity Attacks}
This class covers threats where a malicious actor gains unauthorized access to
artifacts or system components. AIBoMGen uses Azure OAuth for secure
\acrshort{API} calls. If authentication is compromised, attackers could gain
access. The system does not yet have alerting for suspicious activity.
Responsibility for such attacks lies with the authentication provider.

\paragraph{Out-of-Scope Platform Vulnerabilities}
Finally, it is important to acknowledge that some threats remain out of scope
for this evaluation. These include unknown or untested platform vulnerabilities
such as zero-day exploits, backdoors, and authentication flaws. While AIBoMGen
implements a range of security controls, it does not claim to defend against
all possible attack vectors, especially those that are not explicitly addressed
or tested in the current implementation.

As mentioned in Section \ref{subsec:architecture}, AIBoMGen also includes a
dedicated vulnerability scanner container, which regularly scans the Docker
images used by the workers for known vulnerabilities. This adds an additional
layer of defense by helping to identify and address security issues in the
container environment before they can be exploited.

\section{Conclusion}

This research investigated how the concept of an \acrshort{AIBOM} can be
extended and implemented as a verifiable approach to ensure transparency in
\acrshort{AI} development, illustrated by the open-source platform AIBoMGen.
The findings address three key research questions:

\textbf{RQ1:} How can trust be effectively integrated into the \acrshort{AI} lifecycle
to ensure transparency, integrity, and ethical concerns?

The \acrshort{PoC} shows that a trusted observer, the platform, can
independently generate the \acrshort{AIBOM} with cryptographic attestations,
integrating trust without expensive hardware like \acrshort{TEE}s. A key lesson
is that many industry-standard training workflows rely on custom Docker
containers and training scripts, which are not inherently secure or attested.
This means there has to be some sort of restriction on the training environment
through a controlled pipeline, making sure that the platform can externally
observe and attest to the process, without being bypassed or manipulated. This
means there is a clear trade-off between the amount of guarantees and the
flexibility of the developer. Concretely, this helps address ethical concerns
such as data consent and licensing and the presence of harmful content, because
these aspects can be recorded as explicit, verifiable fields in the
\acrshort{AIBOM} instead of only in informal documentation. Since datasets and
training configurations are captured in a structured way, integrating external
bias or fairness analysis tools on top of generated \acrshort{AIBOM}s becomes
straightforward.

\textbf{RQ2:} To what extent can trust mechanisms be integrated without
excessively limiting developer flexibility and innovation?

AIBoMGen shows that strong trust mechanisms such as enforced pipelines, and
cryptographic proofs can be added with little performance or storage cost. The
downside is that these safeguards limit developer freedom. You cannot run
arbitrary code, and all training must stay inside a controlled workflow. This
ensures integrity and verifiability, but makes it harder to use custom scripts,
new frameworks, or unusual workflows unless they are explicitly integrated into
the platform. In the end, it is a trade-off: stronger trust and security
usually mean less flexibility. Careful design, such as modular pipelines or
future use of trusted execution environments, can help restore flexibility as
security technology matures.

\textbf{RQ3:} Which concrete methods and technologies enable the automated generation
of \acrshort{AIBOM}s while maintaining their trustworthiness?

The \acrshort{PoC} shows that automation is feasible using tools like CycloneDX
Python library, in-toto (or other attestation tools), and containerized
workflows. The reliance on emerging standards and incomplete library support
(e.g., CycloneDX Python library) also underscores the need for further
development in this area. \acrshort{AIBOM}s can play an important role in
meeting transparency requirements of the \acrshort{AI} Act. It provides a
verifiable record of datasets, dependencies, and training processes, which
enables organizations to demonstrate compliance. However, the lack of
widespread adoption, and most of all, tooling for \acrshort{AIBOM}s remains a
barrier.

This research highlights the potential of AIBoMGen (and similar platforms) as a
foundational step toward building secure and transparent \acrshort{AI}
ecosystems, enabling compliance with regulatory frameworks like the
\acrshort{EU} \acrshort{AI} Act.

\section*{Future Work}
Future improvements include integrating \acrshort{TEE} support to enable hardware-rooted
attestations and potentially allow for more user-written scripts inside trusted environments. While
prior work has explored attestations using \acrshort{TEE}s, further research is
needed to adapt these techniques for broader applicability and to support more flexible user
workflows. Another area for enhancement is the adoption of modern signing and transparency tools,
such as Sigstore and Rekor, for secure management of \acrshort{AIBOM}s. Extending support to
additional frameworks and pipelines, for example through integration with platforms like Kubeflow
Pipelines, would also increase the system’s versatility. There should be
a mechanism to automatically link different \acrshort{AIBOM}s and provide a full lineage
chain. This has significant potential for improving auditing and provenance tracking.

Another promising direction is to move beyond secure model file formats ("safe
tensors") and investigate the concept of "safe trainers." This would involve
fundamental changes to popular \acrshort{AI} frameworks to enable them to
produce verifiable evidence of the training process itself. For example,
frameworks could be extended to provide standardized \acrshort{API}s for
external attestation tools to monitor and verify the training workflow. Such
modifications would make it possible to prove that a model was trained as
claimed, without hidden manipulations or bypasses. Achieving this level of
trustworthiness would require close collaboration between framework developers,
the research community, and industry stakeholders.

\begin{acks}
    This work has been partially supported by the CRACY project, funded by the European Union’s Digital Europe Programme under grant agreement No 101190492.
\end{acks}

\bibliographystyle{ACM-Reference-Format}
\bibliography{references}

@misc{marr_examples_2024,
  title    = {Examples {That} {Illustrate} {Why} {Transparency} {Is} {Crucial} {In} {AI}},
  url      = {https://www.forbes.com/sites/bernardmarr/2024/05/17/examples-that-illustrate-why-transparency-is-crucial-in-ai/},
  language = {en},
  urldate  = {2026-01-05},
  journal  = {Forbes},
  author   = {Marr, Bernard},
  month    = may,
  year     = {2024},
  note     = {Section: Enterprise Tech}
}

@misc{ghosh_ai_2024,
  title    = {{AI} {Bill} of {Materials} ({AI} {BOM})},
  url      = {https://medium.com/@bijit211987/ai-bill-of-materials-ai-bom-80d48f9d75e0},
  urldate  = {2024-10-28},
  author   = {Ghosh, Bijit},
  month    = jul,
  year     = {2024}
}

@misc{vandendriessche_aibomgen-experiments_2025,
  title     = {{AIBoMGen}-experiments {Scripts} \& {Results}},
  url       = {https://zenodo.org/records/15505280},
  urldate   = {2025-05-28},
  publisher = {Zenodo},
  author    = {Vandendriessche, Wiebe},
  month     = may,
  year      = {2025},
  doi       = {10.5281/zenodo.15505280}
}

@misc{vandendriessche_aibomgen_2025,
  title     = {{AIBoMGen}},
  url       = {https://zenodo.org/records/15536532},
  urldate   = {2025-05-28},
  publisher = {Zenodo},
  author    = {Vandendriessche, Wiebe},
  month     = may,
  year      = {2025},
  doi       = {10.5281/zenodo.15536532}
}

@misc{minio_minio_2025,
  title    = {{MinIO} - {S3} {Compatible} {Storage} for {AI}},
  url      = {https://min.io},
  urldate  = {2025-05-17},
  author   = {{MinIO}},
  year     = {2025}
}

@misc{owasp_foundation_cyclonedx_2024,
  title    = {{CycloneDX} - {Software} {Bill} of {Materials} ({SBOM})},
  url      = {https://cyclonedx.org/},
  urldate  = {2024-10-22},
  author   = {{OWASP Foundation}},
  month    = jul,
  year     = {2024}
}

@misc{owasp_foundation_cyclonedx_2024-1,
  title    = {{CycloneDX} - {Machine} {Learning} {Bill} of {Materials} ({ML}-{BOM})},
  url      = {https://cyclonedx.org/capabilities/mlbom/},
  urldate  = {2024-10-22},
  author   = {{OWASP Foundation}},
  year     = {2024}
}

@misc{linux_foundation_spdx_2024,
  title    = {{SPDX} {AI} {Github} {Specification} 3.0.1},
  url      = {https://spdx.github.io/spdx-spec/v3.0.1/model/AI/AI/},
  urldate  = {2024-10-22},
  author   = {{Linux Foundation}},
  year     = {2024}
}

@misc{linux_foundation_spdx_2023,
  title    = {{SPDX}},
  url      = {https://spdx.dev/},
  urldate  = {2024-10-21},
  author   = {{Linux Foundation}},
  year     = {2023}
}

@misc{bardenstein_whitepaper_2023,
  title    = {Whitepaper: {Driving} {AI} {Transparancy}: the {AI} {Bill} of {Materials} ({Manifest})},
  url      = {manifestcyber.com},
  urldate  = {2024-10-29},
  author   = {Bardenstein, Daniel},
  year     = {2023}
}

@misc{techworks_trustworthy_2023,
  title    = {Trustworthy {AI} - {Practical} {Collaborative} {Engineering}},
  url      = {https://www.techworks.org.uk/wp-content/uploads/2024/01/Engineering-Trustworthy-AI.pdf},
  urldate  = {2025-05-18},
  author   = {{TechWorks} and {NquiringMinds}},
  year     = {2023}
}

@misc{rabbitmq_rabbitmq_2025,
  title      = {{RabbitMQ}: {One} broker to queue them all},
  shorttitle = {{RabbitMQ}},
  url        = {https://www.rabbitmq.com/},
  urldate    = {2025-05-20},
  author     = {{RabbitMQ}},
  year       = {2025}
}

@misc{in-toto_-toto_2025,
  title    = {in-toto 3.0.0 documentation},
  url      = {https://in-toto.readthedocs.io/en/latest/},
  urldate  = {2025-05-18},
  author   = {{in-toto}},
  year     = {2025}
}

@misc{in-toto_-toto_2025-1,
  title    = {in-toto},
  url      = {https://in-toto.io/},
  urldate  = {2025-05-18},
  author   = {{in-toto}},
  year     = {2025}
}

@misc{mithril_security_aicert_2025,
  title    = {{AICert}},
  url      = {https://aicert.mithrilsecurity.io/en/latest/},
  urldate  = {2025-05-18},
  author   = {{Mithril Security}},
  year     = {2025}
}

@misc{hugging_face_hugging_2025,
  title    = {Hugging {Face}},
  url      = {https://huggingface.co/},
  urldate  = {2025-05-18},
  author   = {{Hugging Face}},
  month    = may,
  year     = {2025}
}

@misc{aqua_security_trivy_2025,
  title    = {Trivy},
  url      = {https://trivy.dev/v0.62/},
  urldate  = {2025-05-17},
  author   = {{Aqua Security}},
  year     = {2025}
}

@misc{solem_celery_2023,
  title    = {Celery 5.5.2 documentation},
  url      = {https://docs.celeryq.dev/en/stable/},
  urldate  = {2025-05-17},
  author   = {Solem, Ask and {contributors}},
  year     = {2023}
}

@misc{oracle_mysql_2025,
  title    = {{MySQL}},
  url      = {https://www.mysql.com/},
  urldate  = {2025-05-17},
  author   = {{Oracle}},
  year     = {2025}
}

@misc{fastapi_fastapi_2025,
  title    = {{FastAPI}},
  url      = {https://fastapi.tiangolo.com/},
  urldate  = {2025-05-17},
  author   = {{FastAPI}},
  year     = {2025}
}

@misc{thiel_investigation_2023,
  title    = {Investigation {Finds} {AI} {Image} {Generation} {Models} {Trained} on {Child} {Abuse}},
  url      = {https://cyber.fsi.stanford.edu/news/investigation-finds-ai-image-generation-models-trained-child-abuse},
  urldate  = {2025-02-17},
  author   = {Thiel, David},
  month    = dec,
  year     = {2023}
}

@misc{ozoani_model_2022,
  title    = {Model {Cards}},
  url      = {https://huggingface.co/blog/model-cards},
  urldate  = {2024-12-16},
  author   = {Ozoani, Ezi and GerChick, Marissa and Mitchell, Margaret},
  year     = {2022}
}

@misc{linux_foundation_implementing_2024,
  title    = {Implementing {AI} {Bill} of {Materials} ({AI} {BOM}) with {SPDX} 3.0},
  url      = {https://www.linuxfoundation.org/research/ai-bom},
  urldate  = {2024-12-14},
  author   = {{Linux Foundation}},
  year     = {2024}
}

@misc{imec_gpulab_2025,
  title    = {{GPULab} - imec {iLab}.t documentation},
  url      = {https://doc.ilabt.imec.be/ilabt/gpulab/},
  urldate  = {2024-11-18},
  author   = {{imec}},
  year     = {2025}
}

@misc{techworks_compliance_2024,
  title    = {Compliance with the {EU} {AI} {Act} {The} {Techworks} {Trusted} {AI} {Bill} {Of} {Materials} ({TAIBOM}) project},
  url      = {https://www.techworks.org.uk/ai-taibom/compliance-with-the-eu-ai-act-the-techworks-trusted-ai-bill-of-materials-taibom-project},
  urldate  = {2024-12-14},
  author   = {{TechWorks}},
  month    = jun,
  year     = {2024}
}

@misc{eu_ai_2024,
  title    = {{AI} {Act} - {Shaping} {Europe}’s digital future},
  url      = {https://digital-strategy.ec.europa.eu/en/policies/regulatory-framework-ai},
  urldate  = {2024-11-11},
  author   = {{EU}},
  month    = sep,
  year     = {2024}
}

@article{kale_provenance_2023,
  title      = {Provenance documentation to enable explainable and trustworthy {AI}: {A} literature review},
  volume     = {5},
  issn       = {2641-435X},
  shorttitle = {Provenance documentation to enable explainable and trustworthy {AI}},
  url        = {https://www.sciengine.com/doi/10.1162/dint_a_00119},
  doi        = {10.1162/dint_a_00119},
  number     = {1},
  urldate    = {2025-07-02},
  journal    = {Data Intelligence},
  author     = {Kale, Amruta and Nguyen, Tin and Harris, Frederick C. and Li, Chenhao and Zhang, Jiyin and Ma, Xiaogang},
  month      = mar,
  year       = {2023},
  pages      = {139--162}
}

@article{longpre_position_2024,
  title    = {Position: {Data} {Authenticity}, {Consent}, \& {Provenance} for {AI} are all broken:what will it take to fix them?},
  number   = {1328},
  journal  = {ICML'24: Proceedings of the 41st International Conference on Machine Learning},
  author   = {Longpre, Shayne and Mahari, Robert and Obeng-Marnu, Naana and Brannon, William and South, Tobin and Gero, Katy and Pentland, Sandy and Kabbara, Jad},
  month    = jul,
  year     = {2024},
  pages    = {32711 -- 32725}
}

@inproceedings{davino_aloha_2025,
  address    = {New York, NY, USA},
  series     = {{EASE} '25},
  title      = {{ALOHA}: {A}({IBoM}) {tooL} {generatOr} for {Hugging} {fAce}},
  isbn       = {979-8-4007-1385-9},
  shorttitle = {{ALOHA}},
  url        = {https://dl.acm.org/doi/10.1145/3756681.3756998},
  doi        = {10.1145/3756681.3756998},
  urldate    = {2026-01-07},
  booktitle  = {Proceedings of the 29th {International} {Conference} on {Evaluation} and {Assessment} in {Software} {Engineering}},
  publisher  = {Association for Computing Machinery},
  author     = {D'Avino, Riccardo and Nocera, Sabato and Bifolco, Daniele and Pepe, Federica and Di Penta, Massimiliano and Scanniello, Giuseppe},
  month      = dec,
  year       = {2025},
  pages      = {929--937}
}

@inproceedings{duddu_laminator_2025,
  address    = {New York, NY, USA},
  series     = {{CODASPY} '25},
  title      = {Laminator: {Verifiable} {ML} {Property} {Cards} using {Hardware}-assisted {Attestations}},
  isbn       = {979-8-4007-1476-4},
  shorttitle = {Laminator},
  url        = {https://dl.acm.org/doi/10.1145/3714393.3726492},
  doi        = {10.1145/3714393.3726492},
  urldate    = {2025-07-02},
  booktitle  = {Proceedings of the {Fifteenth} {ACM} {Conference} on {Data} and {Application} {Security} and {Privacy}},
  publisher  = {Association for Computing Machinery},
  author     = {Duddu, Vasisht and Gunn, Lachlan J. and Asokan, N.},
  month      = jun,
  year       = {2025},
  pages      = {317--328}
}

@inproceedings{stalnaker_boms_2024,
  address    = {Lisbon Portugal},
  title      = {{BOMs} {Away}! {Inside} the {Minds} of {Stakeholders}: {A} {Comprehensive} {Study} of {Bills} of {Materials} for {Software} {Systems}},
  isbn       = {979-8-4007-0217-4},
  shorttitle = {{BOMs} {Away}! {Inside} the {Minds} of {Stakeholders}},
  url        = {https://dl.acm.org/doi/10.1145/3597503.3623347},
  doi        = {10.1145/3597503.3623347},
  urldate    = {2025-07-02},
  booktitle  = {Proceedings of the {IEEE}/{ACM} 46th {International} {Conference} on {Software} {Engineering}},
  publisher  = {ACM},
  author     = {Stalnaker, Trevor and Wintersgill, Nathan and Chaparro, Oscar and Di Penta, Massimiliano and German, Daniel M and Poshyvanyk, Denys},
  month      = feb,
  year       = {2024},
  pages      = {1--13}
}

@article{balasubramaniam_transparency_2023,
  title      = {Transparency and explainability of {AI} systems: {From} ethical guidelines to requirements},
  volume     = {159},
  issn       = {0950-5849},
  shorttitle = {Transparency and explainability of {AI} systems},
  url        = {https://www.sciencedirect.com/science/article/pii/S0950584923000514},
  doi        = {10.1016/j.infsof.2023.107197},
  urldate    = {2025-07-03},
  journal    = {Information and Software Technology},
  author     = {Balasubramaniam, Nagadivya and Kauppinen, Marjo and Rannisto, Antti and Hiekkanen, Kari and Kujala, Sari},
  month      = jul,
  year       = {2023},
  pages      = {107197}
}

@article{jobin_global_2019,
  title    = {The global landscape of {AI} ethics guidelines},
  volume   = {1},
  doi      = {10.1038/s42256-019-0088-2},
  journal  = {Nature Machine Intelligence},
  author   = {Jobin, Anna and Ienca, Marcello and Vayena, Effy},
  month    = sep,
  year     = {2019},
  pages    = {389--399}
}

@inproceedings{tomer_experimenting_2022,
  address   = {Chicago, IL, USA},
  title     = {Experimenting an {Edge}-{Cloud} {Computing} {Model} on the {GPULab} {Fed4Fire} {Testbed}},
  url       = {https://ieeexplore.ieee.org/abstract/document/9820006},
  doi       = {10.1109/LANMAN54755.2022.9820006},
  urldate   = {2026-01-07},
  booktitle = {2022 {IEEE} {International} {Symposium} on {Local} and {Metropolitan} {Area} {Networks} ({LANMAN})},
  author    = {Tomer, Vikas and Sharma, Sachin},
  publisher = {IEEE Computer Society},
  month     = jul,
  year      = {2022},
  note      = {ISSN: 1944-0375},
  pages     = {1--2}
}

@inproceedings{spoczynski_atlas_2025,
  title      = {Atlas: {A} {Framework} for {ML} {Lifecycle} {Provenance} \& {Transparency}},
  isbn       = {979-8-3315-9546-3},
  shorttitle = {Atlas},
  url        = {https://www.computer.org/csdl/proceedings-article/euros-pw/2025/954600a448/29EHFNI5Cda},
  doi        = {10.1109/EuroSPW67616.2025.00058},
  urldate    = {2025-09-09},
  booktitle  = {2025 {IEEE} {European} {Symposium} on {Security} and {Privacy} ({EuroS}\&{P} 2025) {Proceedings}},
  publisher  = {IEEE Computer Society},
  author     = {Spoczynski, Marcin and Melara, Marcela S. and Szyller, Sebastian},
  month      = jun,
  address    = {Venice, Italy},
  year       = {2025},
  pages      = {448--461}
}

@misc{tensorflow_tfkerasmodelsload_model_2024,
  title    = {tf.keras.models.load\_model},
  url      = {https://www.tensorflow.org/api_docs/python/tf/keras/models/load_model},
  urldate  = {2025-05-17},
  author   = {{TensorFlow}},
  month    = jun,
  year     = {2024}
}

@misc{righi_tensorflow_2022,
  type     = {{TensorFlow} {Remote} {Code} {Execution} with {Malicious} {Model}},
  title    = {{TensorFlow} {Remote} {Code} {Execution} with {Malicious} {Model}},
  url      = {https://splint.gitbook.io/cyberblog/security-research/tensorflow-remote-code-execution-with-malicious-model},
  urldate  = {2025-03-23},
  author   = {Righi, Tobia},
  month    = oct,
  year     = {2022}
}

\end{document}